\documentclass[%
reprint,
superscriptaddress,
amsmath,amssymb,
aps,
prl,
longbibliography,
lengthcheck,%
]{revtex4}

\usepackage{graphicx}
\usepackage{dcolumn}
\usepackage{bm}
\usepackage{hyperref}

\usepackage[usenames]{color}

\newcommand{\omx}{\omega_x}
\newcommand{\omy}{\omega_y}
\newcommand{\omz}{\omega_z}
\newcommand{\dom}{\Delta}  
\newcommand{\domo}{\Delta_0}
\newcommand{\dR}{\Delta_R}
\newcommand{\lcRate}{\gamma_{c}}
\newcommand{\isreRate}{\omega_{\textit{ex}}}
\newcommand{\Nat}{N_{\text{at}}}
\newcommand{\Bmag}{B_m}
\newcommand{\ket}[1]{\left |#1\right\rangle}

\begin{document}

\preprint{APS/123-QED}

\title{Spin self-rephasing and very long coherence times
 in a trapped atomic ensemble}

\author{C. Deutsch}
\affiliation{Laboratoire Kastler Brossel, ENS, UPMC, CNRS, 24 rue Lhomond, 75005 Paris, France}

\author{F. Ramirez-Martinez}
\affiliation{LNE-SYRTE, Observatoire de Paris, UPMC, CNRS, 61 av de l'Observatoire, 75014 Paris, France}%
\author{C. Lacro\^ute}
\affiliation{LNE-SYRTE, Observatoire de Paris, UPMC, CNRS, 61 av de l'Observatoire, 75014 Paris, France}%
\author{F. Reinhard}
\altaffiliation[Current address: ]{3. Physikal. Institut, Universit\"at Stuttgart, Germany}%
\affiliation{Laboratoire Kastler Brossel, ENS, UPMC, CNRS, 24 rue Lhomond, 75005 Paris, France}%
\author{T. Schneider}
\altaffiliation[Current address: ]{Heinrich-Heine Universit\"at D\"usseldorf, Germany}%
\affiliation{Laboratoire Kastler Brossel, ENS, UPMC, CNRS, 24 rue Lhomond, 75005 Paris, France}%

\author{J. N. Fuchs}
\affiliation{Laboratoire de Physique des Solides, CNRS UMR 8502, Univ. Paris-Sud, 91405 Orsay, France}%
\author{F. Pi\'echon}
\affiliation{Laboratoire de Physique des Solides, CNRS UMR 8502, Univ. Paris-Sud, 91405 Orsay, France}%

\author{F. Lalo\"e}
\affiliation{Laboratoire Kastler Brossel, ENS, UPMC, CNRS, 24 rue Lhomond, 75005 Paris, France}

\author{J. Reichel}
\affiliation{Laboratoire Kastler Brossel, ENS, UPMC, CNRS, 24 rue Lhomond, 75005 Paris, France}

\author{P. Rosenbusch}%
\email{Peter.Rosenbusch@obspm.fr}
\affiliation{LNE-SYRTE, Observatoire de Paris, UPMC, CNRS, 61 av de l'Observatoire, 75014 Paris, France}

\date{\today}

\begin{abstract}
 We perform Ramsey spectroscopy on the ground state of ultra-cold
 $^{87}$Rb atoms magnetically trapped on a chip in the Knudsen
 regime. Field inhomogeneities over the sample should limit the
 $1/e$ contrast decay time to about $3$\,s,  while decay times of $58\pm12$\,s are actually
 observed. We explain this surprising result by a spin self-rephasing
 mechanism induced by the identical spin rotation effect
 originating from particle indistinguishability. We propose a
 theory of this synchronization mechanism and obtain good agreement
 with the experimental observations. The effect is general and
may appear in other physical systems.

\end{abstract}

\pacs{Valid PACS appear here}

\maketitle

In atomic clocks and other precision techniques based on atomic spin
manipulation \cite{Chu02}, a central requirement is to preserve the
coherence of a state superposition over long times. Understanding how
coherence decays in a given system is important for these
applications, and is a touchstone of understanding its dynamics. In
trapped ensembles, an inhomogeneous shift $\dom(\mathbf{r})$ of the
transition frequency occurs due to the trapping potential and to
atomic interactions. Different atoms explore different regions of this
shift landscape, and so their spins precess at different rates. This
leads to dephasing at a rate determined by the characteristic
inhomogeneity $\domo$ of $\dom(\mathbf{r})$ over the ensemble. Various
mechanisms have been exploited to reduce this dephasing. Examples are
``magic fields''  that strongly reduce the field dependence for
a specific transition \cite{Katori03,Harber02}, or the mutual
compensation scheme successfully employed in ultracold $^{87}$Rb
\cite{Lewandowski02}, where the trap-induced inhomogeneity can be
adjusted to nearly cancel the collisional mean-field
inhomogeneity. All such mechanisms however, including the motional
narrowing well known in nuclear magnetic resonance, have in common
that the dephasing is merely slowed down, but never reversed, and the
transverse polarization remains a steadily decreasing function of
time.

\begin{figure}[t]
\includegraphics[width= 0.99 \columnwidth]{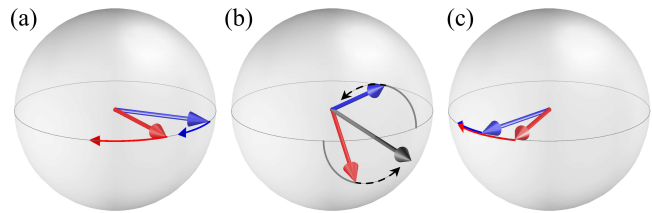}
\caption{Two classes of atoms (red and blue) precess at different
 rates. Their Bloch vectors were initially parallel, but have
 started to dephase (a). The ISRE then makes both vectors rotate
 around their sum (grey) (b). When this rotation reaches $\pi$, the
 fast-precessing spin (red) lags behind the slow one (blue), which
 tends to rephase them as in a spin echo (c).}
\label{fig:Bloch}
\end{figure}

Here we present measurements on a trapped ensemble of $^{87}$Rb atoms
with two internal levels
 equivalent to a spin 1/2. Atomic interactions cause a
spontaneous \textit{re-phasing} of the spins, observed as a
 much longer decay time and revivals of Ramsey contrast. We are also
able to extend the coherence time by more than an order of magnitude
beyond the 2 to 3\,s previously achieved on this system
\cite{Harber02,Treutlein04}.
We explain these remarkable results by a very general mechanism based
on the identical spin rotation effect (ISRE) that occurs during
collisions in the forward direction between two identical particles
\cite{Lhuillier82} - an equivalent description can be given in
 terms of the exchange mean-field experienced by the atoms
 \cite{Bashkin81}. This effect is known to cause transient spin
waves
\cite{Lewandowski02,Du08,Du09,Oktel02,Fuchs02,Williams02,Piechon09,Natu09},
a deleterious phenomenon if one is interested in long coherence times.
In contrast to those experiments however, we are working in a regime
where both (i) the ISRE rate (exchange rate)
$\isreRate/2\pi=2\hbar|a_{01}|\bar{n}/m$ is larger than the
inhomogeneity $\domo$ {\it and} (ii) the rate of lateral elastic
collisions $\lcRate=(32\sqrt{\pi}/3)a_{01}^2\bar{n}v_T$ is much lower
than both the trap frequencies (Knudsen regime) and
$\isreRate$. Here, $a_{01}$ is the relevant scattering length
\footnote{This value of $\gamma_c$ assumes $a_{00}\approx
 a_{11}\approx a_{01}$, as in our experiment.}, $\bar{n}$ the
average density and $v_T\equiv\sqrt{k_B T/m}$ the thermal velocity for
atomic mass $m$.  In this regime, we find that the ISRE introduces
efficient synchronization between atoms with different spin precession
rates.
Indeed, solving a kinetic equation for the spin variables
based on the ISRE, we obtain good agreement with the data. These
findings are reminiscent of earlier calculations for a trapped gas
which predict {\it localized} polarization revivals \cite{Williams02}
and synchronization within spatial domains \cite{Oktel02}  in a different (hydrodynamic) regime.

We start with a simple model (figure~\ref{fig:Bloch}), where as in
 \cite{Piechon09} we divide the atoms into two classes having fast
and slow transverse spin precession rates. Whether an atom
is in the fast or slow class depends on the average of
$\dom(\mathbf{r})$ it experiences. In the second experiment
below, for example, this depends on its orbital energy. If
 the class-changing events are rare ($\lcRate<\isreRate$), an atom
remains in its class for a long time and the two classes start to
dephase. The effect of the ISRE is then simply to make the spin
polarizations of the slow and fast class turn around their sum,
reversing their phase differences on the timescale of
$\pi/\isreRate$. When this rotation reaches $\pi$, the two
transverse polarizations are exchanged, while each of them continues
to precess at the same rate as before, so that they start to
rephase.

The ISRE does not change the sum of the polarizations:
it just reverts the correlation between the spin directions and their
precession rate. This initiates the rephasing process that
increases the transverse polarization. Of course, in reality dephasing
and refocussing occur simultaneously; the result is a synchronization
mechanism that is analogous to a negative feedback effect. It is also
reminiscent of a spin echo, but here the rephasing of the spins is
due to an internal effect instead of externally applied RF pulses.

For a self-rephasing regime to occur, the ISRE must revert the spins before lateral collisions cause
atoms to change class -- that is, $\isreRate /\pi>\lcRate$  --
 and before the dephasing reaches $\pi$, $\isreRate>\domo$ (otherwise it accelerates the dephasing). If $\isreRate\gg\domo$, any dephasing will be immediately refocussed, leading to tight synchronization and long coherence
times; if $\isreRate \gtrsim \domo$, a significant phase spread
can occur before the spins rephase, leading to loss and revival of the total polarization.
Performing two experiments, we have indeed observed these behaviours, and found
good agreement with our quantitative  calculations.

We perform Ramsey spectroscopy on the
$\ket{0}\equiv\ket{F=1,m_F=-1}$ to $\ket{1}\equiv\ket{F=2,m_F=1}$
hyperfine transition of $^{87}$Rb atoms magnetically trapped on a
chip.  The differential Zeeman shift $\dom_{Z}(\mathbf{r})$ is
quadratic in magnetic field around a ``magic'' value $\Bmag=
3.228917(3)\,$G \cite{Harber02} and quartic in position. The s-wave
scattering lengths $a_{00}$, $a_{11}$ and $a_{01}$ differ by less than
5\% \cite{vanKempen02}, so that the mean-field shift of the transition
frequency is already small with $\dom_{mf}(\mathbf{r})/ 2\pi=-
0.4~\textrm{Hz}\times n(\mathbf{r})/10^{12}\textrm{cm}^{-3}$
for an equal superposition of $\ket{0}$ and $\ket{1}$, where
$n(\mathbf{r})$ is the local density. Additionally, we employ the
mutual compensation technique \cite{Harber02}: adjusting the magnetic
field at the trap center $B_0$ to a value slightly below $\Bmag$ leads to a
standard deviation of
$\dom(\mathbf{r})\equiv\dom_Z(\mathbf{r})+\dom_{mf}(\mathbf{r})$
averaged over the atomic cloud that is of order
$\domo=2\pi\times0.08$\,Hz \cite{Rosenbusch09}.

The experiment is designed with the goal of a ``trapped-atom clock on
a chip'' (TACC), aiming at a stability in the lower
$10^{-13}\mathrm{s}^{-1/2}$ range.  The setup is described in detail
in \cite{Lacroute10}, and is similar to that of \cite{Boehi09}. It
incorporates a two-layer atom chip with a coplanar waveguide
(CPW). Atom preparation involves the usual laser and evaporative
cooling steps \cite{Haensel01a}. Within less than 10\,s, we obtain
$\Nat=5\times10^3$ to $10^5$ trapped atoms in state $\ket{0}$ at a
temperature of $T=175(6)\,$nK, which is at least 30\,nK above the
onset of Bose Einstein condensation (BEC) for our densities. We
use a two-photon, microwave (MW) and radiofrequency (RF) excitation to
drive the clock transition. The MW frequency is detuned $\sim
500$\,kHz above the $\ket{0}$ to $\ket{F=2,m_F=0}$ transition. The
detuning from the two-photon resonance is set to
$\dR/2\pi=3.6$~Hz. The MW is generated by a home-built
synthesizer having very low phase noise \cite{Ramirez10}; a commercial
direct digital synthesizer provides the RF. Both are locked to
a hydrogen maser of relative frequency stability $10^{-13}
\tau^{-1/2}$ up to 1000\,s \cite{Chambon05}.  Both signals are
injected into structures on the chip with powers $\sim 0$\,dBm (MW)
$\sim$5\,dBm (RF).  Trapped below the CPW at a distance $z_0$, the
atoms interact with its evanescent field and perform two-photon Rabi
oscillations at a rate $\Omega_R$.
In each cycle, we detect the number of atoms in both clock states,
$N_1$ and $N_0$, by absorption imaging without and with repump light
after a time of flight of 9\,ms and 13\,ms respectively.  We use
absorption imaging close to saturation intensity \cite{Reinaudi07}.
Intensity and magnification are carefully calibrated, and we have
verified that cross talk between the detected $N_0$ and $N_1$ is
negligible. We estimate the calibration error on the absolute number
of atoms to 5\%.  The statistical error is $140$ atoms per state.

\begin{figure*}[t]
\includegraphics[width=1.\textwidth]{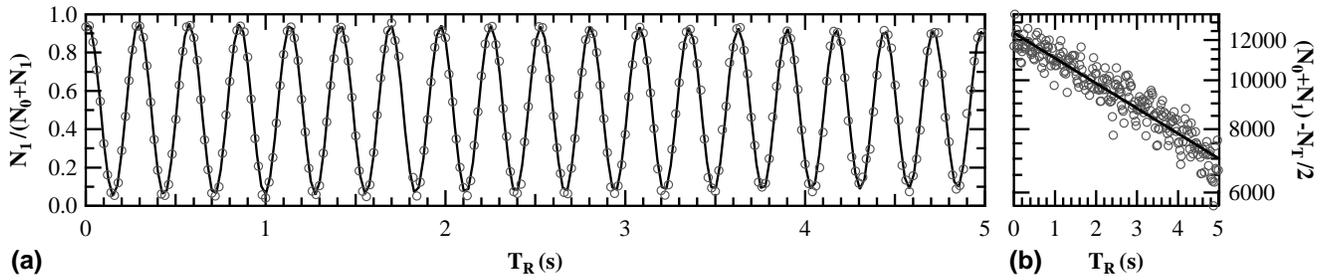}
\caption{First experiment: (a) Ramsey fringes in the time domain for
 $\domo/2\pi\approx0.08$\,Hz. The normalized transition probablity is
 plotted as a function of the Ramsey free evolution time. The initial
 contrast is 89\%, probably limited by an inhomogeneous Rabi
 frequency. At $T_R=5$\,s, the contrast is 82\% resulting in a 1/e
 time of $58\pm12$~s. (b) Total number of detected atoms. Although
 the total time $T_T$ is kept constant, this number depends on $T_R$
 because the loss rates of $\ket{0}$ and $\ket{1}$ differ. The data
 is fitted by $N_0+N_1= N_{T}/2\; (1+e^{-T_R/\tau})$ with
   $N_{T}=24.8(2)\times10^3$ atoms and
   $\tau=8.7(2)$~s.\label{fig:coherence}}
 \label{Fig-long-coherence}
\end{figure*}

In a first experiment, the magnetic trap has frequencies
$\{\omx,\omy,\omz\}/2\pi= \{32(1),97.5(2.5),121(1)\}$\,Hz; for this trap $z_0=156\,\mu$m and $\Omega_R=2\pi\times 164\,$Hz.
$B_0$ is optimized roughly by maximizing the fringe contrast after a
Ramsey time $T_R=2\,$s. This leads to $B_0=3.1626(7)$\,G, as measured
by RF-induced atom loss on a small BEC ($\Nat\sim 3000$).
The two $\pi/2$ pulses, spaced by a variable Ramsey time
$T_R$, are applied at the end of a period of fixed length of $T_T=5.02\,$s, and preceded by
a holding time $T_H=T_T-T_R$ during which the atoms are in
$\ket{0}$. The transition probability $N_1/(N_0+N_1)$ is
plotted in figure \ref{fig:coherence}(a) as a function of $T_R$.  The
contrast at $T_R=0$ is $89$\%, probably limited by a small
inhomogeneity of the Rabi frequency. At $T_R=5$\,s the contrast still
remains above $82$\%. Assuming exponential decay, the $1/e$ time is
$58\pm 12$\,s,  much longer than
the 2.75\,s predicted from the mutual compensation scheme
\cite{Rosenbusch09}.  The true dephasing is even slower,
since the atom loss rates $\gamma_0$ and $\gamma_1$ of
$\ket{0}$ and $\ket{1}$ differ (fig.~\ref{fig:coherence}(b));
asymmetric loss leads to a decay of Ramsey contrast that is
independent of dephasing.

The very long coherence time can be understood as a tight
synchronization due to self-rephasing. The density $\bar{n}\approx 1$
(from now on in units of $10^{12}\text{cm}^{-3}$) gives an ISRE rate
$\isreRate/2\pi\approx 8$\,Hz, much larger than
the inhomogeneity $\domo/2\pi\approx 0.08$\,Hz, so that the fast and
slow spins are swapped when the dephasing is still small.
Moreover, since $\lcRate\approx2\,\textrm{s}^{-1}$,
the correlations between precession rate
and accumulated dephasing remain intact over long times.

\begin{figure}[b]
\includegraphics[width= 1.0 \columnwidth]{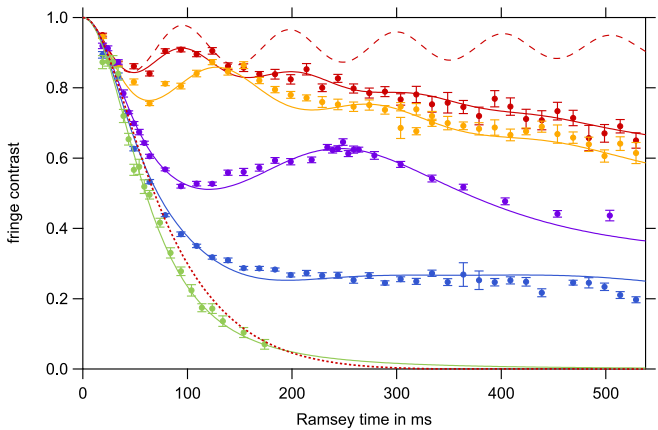}
 \caption{\label{fig:contrast} Second experiment: Ramsey fringe contrast as a function of $T_R$ for increased inhomogeneity
   $\domo/2\pi\approx 1.2\,$Hz. From bottom to top, colors correspond to densities
   $\bar{n}=\{0.2,0.8,1.1,1.9,2.6\}\times 10^{12}$
   cm$^{-3}$
   (or $N_{at}=\{7.9,28,41,70,95\}\times10^3$).
   For each $\bar{n}$, the data is normalized to the initial contrast at
   $T_R=0$. Solid lines are numerical solutions of the kinetic
   equation for $\domo/2\pi\approx 2$ Hz taking into account the
   ISRE ($\omega_{ex}/2\pi \approx 4.5 \textrm{Hz} \times
     \bar{n}$) and lateral collisions ($\gamma_c\approx
     2.1\textrm{s}^{-1} \times \bar{n}$).  The red dotted line is a
   calculation for $\bar{n}=2.6$ (same density as the red data
     points) with lateral collisions ($\gamma_c= 5
   \textrm{s}^{-1}$), but without ISRE.  The red
   dashed line is a calculation without lateral collisions
   but with ISRE ($\isreRate/2\pi=12$\,Hz).}
   \end{figure}

In a second experiment, we deliberately
increase $\domo$ by detuning $B_0$ away from the optimal value.
We choose $B_0=3.7562(9)$\,G, such that the
inhomogeneity of the transverse precession rate is now well
approximated by a parabolic spatial dependence $\dom(\mathbf{r})
\approx \domo \left((x/x_T)^2+(y/y_T)^2+(z/z_T)^2\right)$ with
$\{x_T,y_T,z_T\}\equiv
v_T/\{\omega_x,\omega_y,\omega_z\}$. 
Here, $\domo=
(x_T^2/2)\langle \partial_x^2 \dom \rangle\approx 2\pi\,
(1.2+0.1\times \bar{n}$) Hz by averaging
over the gaussian density profile $n(\mathbf{r})$.
Note that $\domo$ is also linear in temperature, which is
constant in our experiment.

The new $B_0$ leads to trap frequencies
$\{\omega_x,\omega_y,\omega_z\}/2\pi = \{ 31.30(5), 92.0(5),
117.0(3)\}\,$Hz; $z_0=151\,\mu$m and $\Omega_R=2\pi\times 83\,$Hz. The
total time is reduced to $T_T=1$\,s and the Ramsey time is
varied between 15 and 500\,ms. For each $T_R$, we vary the detuning
$\dR$ in 30 steps and extract the fringe contrast. We repeat the
measurement for different $\bar{n}$ spanning one order of
magnitude. To vary $\bar{n}$, we vary the atom number by changing the
MOT loading time. (We have checked that the temperature is independent
of $\bar{n}$ within our measurement precision.) The lowest density is
$\bar{n}=0.2$, which corresponds to $\omega_{ex}/2\pi= 1.5$~Hz and
$\gamma_c= 0.4 \textrm{s}^{-1}$.

The results are plotted in figure \ref{fig:contrast}. All datasets
show initial rapid contrast decay between 0 and 50\,ms. For
$\bar{n}=0.2$, it vanishes completely at
$\sim200$\,ms. Increasing the density stabilizes the contrast beyond
200\,ms and leads to revivals. The contrast improves with
increasing density, indicating an interaction effect. The time of the
first revival depends on density with $T_{\text{revival}}\approx
-0.02\,\text{s}+0.3\,\text{s}/ \bar{n}$. This
agrees to within a factor of 2 with the estimation $2\pi/\isreRate=0.13 \textrm{s}/\bar{n}$ from our simple
model above.

To interpret this data, we now perform a quantitative
calculation. As in \cite{Fuchs02,Williams02}, the motion in the trap
is treated semiclassically, while a full quantum treatment of the
spin variables is included in a kinetic equation for a density
operator $\hat{\rho}(\mathbf{r},\mathbf{p},t)$, characterized
by the usual Bloch vector $\mathbf{S}=\textrm{Tr}(\hat{\rho}
\hat{\boldsymbol\sigma}/2)$. The vector components of $\mathbf{S}$
and the Pauli matrices $\hat{\boldsymbol\sigma}$ point along
the three unit vectors of the Bloch sphere $\{\mathbf{u}_{\perp 1},
\mathbf{u}_{\perp 2}, \mathbf{u}_{\parallel} \}$. To describe
the harmonic oscillator in phase space, instead of the usual
$\{x,p_x\}$ (and similarly for $y$ and $z$), we use energy-angle
variables $\{E_x= (x^2+p_x^2)/2,\alpha_x= \arctan (p_x/x)\}$, where
$E_x$, $x$ and $p_x$ are respectively in units of $k_BT$, $x_T$ and
$mv_T$. As the oscillatory motion is fast compared to the spin
dynamics ($\omega_{x,y,z}\gg\isreRate,\domo,\lcRate$), we average
$\mathbf{S}(\mathbf{E},\boldsymbol{\alpha},t)$ over the angles
$\boldsymbol{\alpha}=(\alpha_x,\alpha_y,\alpha_z)$ and obtain
an equation for a spin density $\mathbf{S}(\mathbf{E},t)$ in energy
space only, where $\mathbf{E}=(E_x,E_y,E_z)$ \footnote{Details
of the derivation will be given elsewhere.}. This description is
inspired by that developed in \cite{Du09} for a quasi-1D gas in the
Knudsen regime, but generalized to 3D and with the effect of
lateral collisions giving rise to a damping term $\propto
\gamma_c$. At this stage, the equation still depends on each $E_i$
separately through the frequency shift $\dom(\mathbf{E})$. However,
since our $\dom(\mathbf{r})$ is parabolic in
$\mathbf{r}$, it becomes $\dom(E)=\domo E$ when averaged over
the angles and only depends on the total energy $E=E_x+E_y+E_z$.
This allows us to write an equation for an energy-isotropic spin
density $\mathbf{S}(E,t)$:
\begin{eqnarray}
 \partial _{t}\mathbf{S}(E,t)+\gamma_c
 \left[
    \mathbf{S}(E,t)-\bar{\mathbf{S}}(t)\right]
 \approx \left[ \dom (E)\mathbf{u}_{\parallel}
\vphantom{ \int_0^\infty}
 \right.\nonumber \\
\left.
+\omega _{ex} \int_0^\infty dE' \frac{E'^2}{2} e^{-E'} K(E,E')
\mathbf{S}(E',t)\right] \times \mathbf{S}(E,t) \label{kees}
\end{eqnarray}
where $\bar{\mathbf{S}}\equiv \int_0^\infty dE
\frac{E^2}{2}e^{-E}\mathbf{S}(E)$ is the average spin and $E^2/2$ is
the 3D harmonic oscillator density of states. Upon angle-averaging,
the spin mean-field, which is local in position, becomes long-ranged
in energy space with a kernel $K(E,E')$. In 1D, this kernel
$K_1(E,E')\approx [\textrm{max}(E,E')|E-E'|]^{-1/4}$ \cite{Du09}. In
3D, it is also long-ranged but cumbersome to use. For simplicity,
when solving (\ref{kees}) numerically, we approximate \footnote{We
checked that in 1D this approximation is very good.} it as
infinite-ranged $K(E,E')\approx 1$. Eq. (\ref{kees}) does not
include atom loss. It contains three experimentally tunable
parameters: the inhomogeneity $\domo(T,\bar{n},...)$, the ISRE rate
$\isreRate\propto |a_{01}| \bar{n}$ and the lateral collision rate
$\gamma_c\propto a_{01}^2\bar{n}\sqrt{T}$.

To compare this theory to the experiment, we first consider the
lowest-density dataset ($\bar{n}=0.2$), where
both the ISRE and lateral collisions are negligible
($\domo\gg \isreRate, \lcRate$). When they vanish,
the kinetic equation is easily solved analytically.  For an initial
condition $\mathbf{S}(E,0)=\mathbf{u}_{\perp 1}$, the contrast is
$|\bar{\mathbf{S}}(t)|=(1+(\domo t)^2)^{-3/2}$. Fitting the experimental
contrast with this result, we find, $\domo/2\pi\approx 2$\,Hz, not far from its expected value $\domo/2\pi\approx
1.2$\,Hz. We then numerically solve eq. (\ref{kees}) for the other
densities $\bar{n}=\{0.8,1.1,1.9,2.6\}$.
Choosing $\domo/2\pi$ within 5\% of the above 2~Hz,
$\gamma_c=(32\sqrt{\pi}/3)a_{01}^2\bar{n}v_T\approx 2.1\textrm{s}^{-1}\times
\bar{n}$ as predicted and $\isreRate/2\pi
\approx 0.6 \times 2\hbar |a_{01}|\bar{n}/m\approx
4.5\textrm{Hz} \times \bar{n}$ reproduces all
data well (solid lines in fig.~\ref{fig:contrast}). The
renormalization of the ISRE rate by a factor 0.6 results from
the overestimation of the synchronization effect through the
infinite-range approximation.

If we set $\isreRate=0$, the theory predicts short
coherence time and no revivals (dotted line in
fig.~\ref{fig:contrast}) for all densities, confirming that the ISRE is responsible for the revivals.
We also solve the $\bar{n}=2.6$ case
without lateral collisions, $\gamma_c=0$: the dashed line
in fig.~\ref{fig:contrast} shows an initial drop and revival, but continues to
oscillate around a constant value. The lateral collisions are therefore responsible for the slow decay of the contrast
at long times.
We note that, in contrast to the mechanism of \cite{Oktel02}, here the spin synchronization
does not result from a simple compensation of the
inhomogeneous longitudinal field $\dom (E)\mathbf{u}_{\parallel }$ by
the exchange mean field $\omega _{ex} \bar{\mathbf{S}}(t)$ as they are
orthogonal. Also, while Ref. \cite{Gibble} discusses the effects of the inhomogeneities
of the probe field during the $\pi/2$ pulses, our work deals with the effects of the static field between them (free precession).

In conclusion, the observed spin self-rephasing and synchronization
are relatively robust when the inhomogeneity to compensate is
not too large: in the first experiment, a large orientation is
still present after more than 10 velocity-changing collisions.  The
effect may occur in any sample where the exchange rate $\isreRate$
is larger than the characteristic dephasing rate $\domo$ and the
lateral collision rate $\lcRate$. The ratio between these rates is
tunable through the temperature dependence of $\isreRate/\lcRate$
and through Feshbach resonances. It should be interesting to
investigate whether the mechanism occurs in systems such as
optical lattice clocks \cite{Katori03}.

\begin{acknowledgments}
 This work was supported from the EURYI award
 ``Integrated Quantum Devices'', the Institut Francilien pour la
 Recherche sur les Atomes Froids (IFRAF), and the Delegation Generale de l'Armement (DGA), contract 07.34.005.
\end{acknowledgments}


\begin{thebibliography}{22}
\expandafter\ifx\csname natexlab\endcsname\relax\def\natexlab#1{#1}\fi
\expandafter\ifx\csname bibnamefont\endcsname\relax
 \def\bibnamefont#1{#1}\fi
\expandafter\ifx\csname bibfnamefont\endcsname\relax
 \def\bibfnamefont#1{#1}\fi
\expandafter\ifx\csname citenamefont\endcsname\relax
 \def\citenamefont#1{#1}\fi
\expandafter\ifx\csname url\endcsname\relax
 \def\url#1{\texttt{#1}}\fi
\expandafter\ifx\csname urlprefix\endcsname\relax\def\urlprefix{URL }\fi
\providecommand{\bibinfo}[2]{#2}
\providecommand{\eprint}[2][]{\url{#2}}

\bibitem[{\citenamefont{Chu}(2002)}]{Chu02}
\bibinfo{author}{\bibfnamefont{S.}~\bibnamefont{Chu}},
 \bibinfo{journal}{Nature} \textbf{\bibinfo{volume}{416}},
 \bibinfo{pages}{206} (\bibinfo{year}{2002}).

\bibitem[{\citenamefont{Katori et~al.}(2003)\citenamefont{Katori, Takamoto,
 Pal'chikov, and Ovsiannikov}}]{Katori03}
\bibinfo{author}{\bibfnamefont{H.}~\bibnamefont{Katori}},
 \bibinfo{author}{\bibfnamefont{M.}~\bibnamefont{Takamoto}},
 \bibinfo{author}{\bibfnamefont{V.~G.} \bibnamefont{Pal'chikov}},
 \bibnamefont{and} \bibinfo{author}{\bibfnamefont{V.~D.}
 \bibnamefont{Ovsiannikov}}, \bibinfo{journal}{Phys. Rev. Lett.}
 \textbf{\bibinfo{volume}{91}}, \bibinfo{pages}{173005}
 (\bibinfo{year}{2003}).

\bibitem[{\citenamefont{Harber et~al.}(2002)\citenamefont{Harber, Lewandowski,
 McGuirk, and Cornell}}]{Harber02}
\bibinfo{author}{\bibfnamefont{D.~M.} \bibnamefont{Harber}},
 \bibinfo{author}{\bibfnamefont{H.~J.} \bibnamefont{Lewandowski}},
 \bibinfo{author}{\bibfnamefont{J.~M.} \bibnamefont{McGuirk}},
 \bibnamefont{and} \bibinfo{author}{\bibfnamefont{E.~A.}
 \bibnamefont{Cornell}}, \bibinfo{journal}{Phys.~Rev.~A}
 \textbf{\bibinfo{volume}{66}}, \bibinfo{pages}{053616}
 (\bibinfo{year}{2002}).

\bibitem[{\citenamefont{Lewandowski et~al.}(2002)\citenamefont{Lewandowski,
 Harber, Whitaker, and Cornell}}]{Lewandowski02}
\bibinfo{author}{\bibfnamefont{H.~J.} \bibnamefont{Lewandowski}},
 \bibinfo{author}{\bibfnamefont{D.~M.} \bibnamefont{Harber}},
 \bibinfo{author}{\bibfnamefont{D.L.}~\bibnamefont{Whitaker}}, \bibnamefont{and}
 \bibinfo{author}{\bibfnamefont{E.~A.} \bibnamefont{Cornell}},
 \bibinfo{journal}{Phys.~Rev.~Lett.} \textbf{\bibinfo{volume}{88}},
 \bibinfo{pages}{070403} (\bibinfo{year}{2002}).

\bibitem[{\citenamefont{Treutlein et~al.}(2004)\citenamefont{Treutlein,
 Hommelhoff, Steinmetz, H{\"a}nsch, and Reichel}}]{Treutlein04}
\bibinfo{author}{\bibfnamefont{P.}~\bibnamefont{Treutlein}},
 \bibinfo{author}{\bibfnamefont{P.}~\bibnamefont{Hommelhoff}},
 \bibinfo{author}{\bibfnamefont{T.}~\bibnamefont{Steinmetz}},
 \bibinfo{author}{\bibfnamefont{T.~W.} \bibnamefont{H{\"a}nsch}},
 \bibnamefont{and} \bibinfo{author}{\bibfnamefont{J.}~\bibnamefont{Reichel}},
 \bibinfo{journal}{Phys.~Rev.~Lett.} \textbf{\bibinfo{volume}{92}},
 \bibinfo{pages}{203005} (\bibinfo{year}{2004}).

\bibitem[{\citenamefont{Lhuillier and Lalo{\"e}}(1982)}]{Lhuillier82}
\bibinfo{author}{\bibfnamefont{C.}~\bibnamefont{Lhuillier}} \bibnamefont{and}
 \bibinfo{author}{\bibfnamefont{F.}~\bibnamefont{Lalo{\"e}}},
 \bibinfo{journal}{J. Phys. (Paris)} \textbf{\bibinfo{volume}{43}},
 \bibinfo{pages}{197 and 225} (\bibinfo{year}{1982}).

\bibitem[{\citenamefont{Bashkin}(1981)}]{Bashkin81}
\bibinfo{author}{\bibfnamefont{E.~P.} \bibnamefont{Bashkin}},
 \bibinfo{journal}{JETP Lett.} \textbf{\bibinfo{volume}{33}},
 \bibinfo{pages}{8} (\bibinfo{year}{1981}).

\bibitem[{\citenamefont{Du et~al.}(2008)\citenamefont{Du, Luo, Clancy, and
 Thomas}}]{Du08}
\bibinfo{author}{\bibfnamefont{X.}~\bibnamefont{Du}},
 \bibinfo{author}{\bibfnamefont{L.}~\bibnamefont{Luo}},
 \bibinfo{author}{\bibfnamefont{B.}~\bibnamefont{Clancy}}, \bibnamefont{and}
 \bibinfo{author}{\bibfnamefont{J.~E.} \bibnamefont{Thomas}},
 \bibinfo{journal}{Phys. Rev. Lett.} \textbf{\bibinfo{volume}{101}},
 \bibinfo{pages}{150401} (\bibinfo{year}{2008}).

\bibitem[{\citenamefont{Du et~al.}(2009)\citenamefont{Du, Zhang, Petricka, and
 Thomas}}]{Du09}
\bibinfo{author}{\bibfnamefont{X.}~\bibnamefont{Du}},
 \bibinfo{author}{\bibfnamefont{Y.}~\bibnamefont{Zhang}},
 \bibinfo{author}{\bibfnamefont{J.}~\bibnamefont{Petricka}}, \bibnamefont{and}
 \bibinfo{author}{\bibfnamefont{J.~E.} \bibnamefont{Thomas}},
 \bibinfo{journal}{Phys. Rev. Lett.} \textbf{\bibinfo{volume}{103}},
 \bibinfo{pages}{010401} (\bibinfo{year}{2009}).

\bibitem[{\citenamefont{Oktel and Levitov}(2002)}]{Oktel02}
\bibinfo{author}{\bibfnamefont{M.~{\"O}.} \bibnamefont{Oktel}}
 \bibnamefont{and} \bibinfo{author}{\bibfnamefont{L.~S.}
 \bibnamefont{Levitov}}, \bibinfo{journal}{Phys.~Rev.~Lett.}
 \textbf{\bibinfo{volume}{88}}, \bibinfo{pages}{230403}
 (\bibinfo{year}{2002}).

\bibitem[{\citenamefont{Fuchs et~al.}(2002)\citenamefont{Fuchs, Gangardt, and
 Lalo\"e}}]{Fuchs02}
\bibinfo{author}{\bibfnamefont{J.~N.} \bibnamefont{Fuchs}},
 \bibinfo{author}{\bibfnamefont{D.~M.} \bibnamefont{Gangardt}},
 \bibnamefont{and} \bibinfo{author}{\bibfnamefont{F.}~\bibnamefont{Lalo\"e}},
 \bibinfo{journal}{Phys. Rev. Lett.} \textbf{\bibinfo{volume}{88}},
 \bibinfo{pages}{230404} (\bibinfo{year}{2002}).

\bibitem[{\citenamefont{Williams et~al.}(2002)\citenamefont{Williams, Nikuni,
 and Clark}}]{Williams02}
\bibinfo{author}{\bibfnamefont{J.~E.} \bibnamefont{Williams}},
 \bibinfo{author}{\bibfnamefont{T.}~\bibnamefont{Nikuni}}, \bibnamefont{and}
 \bibinfo{author}{\bibfnamefont{C.~W.} \bibnamefont{Clark}},
 \bibinfo{journal}{Phys. Rev. Lett.} \textbf{\bibinfo{volume}{88}},
 \bibinfo{pages}{230405} (\bibinfo{year}{2002}).

\bibitem[{\citenamefont{Pi\'echon et~al.}(2009)\citenamefont{Pi\'echon, Fuchs,
 and Lalo\"e}}]{Piechon09}
\bibinfo{author}{\bibfnamefont{F.}~\bibnamefont{Pi\'echon}},
 \bibinfo{author}{\bibfnamefont{J.~N.} \bibnamefont{Fuchs}}, \bibnamefont{and}
 \bibinfo{author}{\bibfnamefont{F.}~\bibnamefont{Lalo\"e}},
 \bibinfo{journal}{Phys. Rev. Lett.} \textbf{\bibinfo{volume}{102}},
 \bibinfo{pages}{215301} (\bibinfo{year}{2009}).

\bibitem[{\citenamefont{Natu and Mueller}(2009)}]{Natu09}
\bibinfo{author}{\bibfnamefont{S.~S.} \bibnamefont{Natu}} \bibnamefont{and}
 \bibinfo{author}{\bibfnamefont{E.~J.} \bibnamefont{Mueller}},
 \bibinfo{journal}{Phys. Rev. A} \textbf{\bibinfo{volume}{79}},
 \bibinfo{pages}{051601(R)} (\bibinfo{year}{2009}).

\bibitem[{\citenamefont{van Kempen et~al.}(2002)\citenamefont{van Kempen,
 Kokkelmans, Heinzen, and Verhaar}}]{vanKempen02}
\bibinfo{author}{\bibfnamefont{E.~G.~M.} \bibnamefont{van Kempen}},
 \bibinfo{author}{\bibfnamefont{S.~J. J. M.~F.} \bibnamefont{Kokkelmans}},
 \bibinfo{author}{\bibfnamefont{D.~J.} \bibnamefont{Heinzen}},
 \bibnamefont{and} \bibinfo{author}{\bibfnamefont{B.~J.}
 \bibnamefont{Verhaar}}, \bibinfo{journal}{Phys.~Rev.~Lett.}
 \textbf{\bibinfo{volume}{88}}, \bibinfo{pages}{093201}
 (\bibinfo{year}{2002}).

\bibitem[{\citenamefont{Rosenbusch}(2009)}]{Rosenbusch09}
\bibinfo{author}{\bibfnamefont{P.}~\bibnamefont{Rosenbusch}},
 \bibinfo{journal}{Applied Physics B: Lasers and Optics}
 \textbf{\bibinfo{volume}{95}}, \bibinfo{pages}{227} (\bibinfo{year}{2009}).

\bibitem[{\citenamefont{Lacroute et~al.}(2010)\citenamefont{Lacroute, Reinhard,
 {Ramirez-Martinez}, Deutsch, Schneider, Reichel, and
 Rosenbusch}}]{Lacroute10}
\bibinfo{author}{\bibfnamefont{C.}~\bibnamefont{Lacroute}},
 \bibinfo{author}{\bibfnamefont{F.}~\bibnamefont{Reinhard}},
 \bibinfo{author}{\bibfnamefont{F.}~\bibnamefont{{Ramirez-Martinez}}},
 \bibinfo{author}{\bibfnamefont{C.}~\bibnamefont{Deutsch}},
 \bibinfo{author}{\bibfnamefont{T.}~\bibnamefont{Schneider}},
 \bibinfo{author}{\bibfnamefont{J.}~\bibnamefont{Reichel}}, \bibnamefont{and}
 \bibinfo{author}{\bibfnamefont{P.}~\bibnamefont{Rosenbusch}},
 \bibinfo{journal}{{IEEE} Transactions on Ultrasonics, Ferroelectrics and
 Frequency Control} \textbf{\bibinfo{volume}{57}}, \bibinfo{pages}{106}
 (\bibinfo{year}{2010}), ISSN \bibinfo{issn}{0885-3010}.

\bibitem[{\citenamefont{B{\"o}hi et~al.}(2009)\citenamefont{B{\"o}hi, Riedel,
 Hoffrogge, Reichel, and H{\"a}nsch}}]{Boehi09}
\bibinfo{author}{\bibfnamefont{P.}~\bibnamefont{B{\"o}hi}},
 \bibinfo{author}{\bibfnamefont{M.~F.} \bibnamefont{Riedel}},
 \bibinfo{author}{\bibfnamefont{J.}~\bibnamefont{Hoffrogge}},
 \bibinfo{author}{\bibfnamefont{J.}~\bibnamefont{Reichel}}, \bibnamefont{and}
 \bibinfo{author}{\bibfnamefont{T.~W.} \bibnamefont{H{\"a}nsch}},
 \bibinfo{journal}{Nature Phys.} \textbf{\bibinfo{volume}{5}},
 \bibinfo{pages}{592} (\bibinfo{year}{2009}).

\bibitem[{\citenamefont{H{\"a}nsel et~al.}(2001)\citenamefont{H{\"a}nsel,
 Hommelhoff, H{\"a}nsch, and Reichel}}]{Haensel01a}
\bibinfo{author}{\bibfnamefont{W.}~\bibnamefont{H{\"a}nsel}},
 \bibinfo{author}{\bibfnamefont{P.}~\bibnamefont{Hommelhoff}},
 \bibinfo{author}{\bibfnamefont{T.~W.} \bibnamefont{H{\"a}nsch}},
 \bibnamefont{and} \bibinfo{author}{\bibfnamefont{J.}~\bibnamefont{Reichel}},
 \bibinfo{journal}{Nature} \textbf{\bibinfo{volume}{413}},
 \bibinfo{pages}{498} (\bibinfo{year}{2001}).

\bibitem[{\citenamefont{{Ramirez-Martinez}
 et~al.}(2010)\citenamefont{{Ramirez-Martinez}, Lours, Rosenbusch, Reinhard,
 and Reichel}}]{Ramirez10}
\bibinfo{author}{\bibfnamefont{F.}~\bibnamefont{{Ramirez-Martinez}}},
 \bibinfo{author}{\bibfnamefont{M.}~\bibnamefont{Lours}},
 \bibinfo{author}{\bibfnamefont{P.}~\bibnamefont{Rosenbusch}},
 \bibinfo{author}{\bibfnamefont{F.}~\bibnamefont{Reinhard}}, \bibnamefont{and}
 \bibinfo{author}{\bibfnamefont{J.}~\bibnamefont{Reichel}},
 \bibinfo{journal}{{IEEE} Transactions on Ultrasonics, Ferroelectrics and
 Frequency Control} \textbf{\bibinfo{volume}{57}}, \bibinfo{pages}{88}
 (\bibinfo{year}{2010}), ISSN \bibinfo{issn}{0885-3010}.

\bibitem[{\citenamefont{Chambon et~al.}(2005)\citenamefont{Chambon, Bize,
 Lours, Narbonneau, Marion, Clairon, Santarelli, Luiten, and
 Tobar}}]{Chambon05}
\bibinfo{author}{\bibfnamefont{D.}~\bibnamefont{Chambon}},
 \bibinfo{author}{\bibfnamefont{S.}~\bibnamefont{Bize}},
 \bibinfo{author}{\bibfnamefont{M.}~\bibnamefont{Lours}},
 \bibinfo{author}{\bibfnamefont{F.}~\bibnamefont{Narbonneau}},
 \bibinfo{author}{\bibfnamefont{H.}~\bibnamefont{Marion}},
 \bibinfo{author}{\bibfnamefont{A.}~\bibnamefont{Clairon}},
 \bibinfo{author}{\bibfnamefont{G.}~\bibnamefont{Santarelli}},
 \bibinfo{author}{\bibfnamefont{A.}~\bibnamefont{Luiten}}, \bibnamefont{and}
 \bibinfo{author}{\bibfnamefont{M.}~\bibnamefont{Tobar}},
 \bibinfo{journal}{Rev. Sci. Instrum.} \textbf{\bibinfo{volume}{76}},
 \bibinfo{pages}{094704} (\bibinfo{year}{2005}).

\bibitem[{\citenamefont{Reinaudi et~al.}(2007)\citenamefont{Reinaudi, Lahaye,
 Wang, and Gu{\'e}ry-Odelin}}]{Reinaudi07}
\bibinfo{author}{\bibfnamefont{G.}~\bibnamefont{Reinaudi}},
 \bibinfo{author}{\bibfnamefont{T.}~\bibnamefont{Lahaye}},
 \bibinfo{author}{\bibfnamefont{Z.}~\bibnamefont{Wang}}, \bibnamefont{and}
 \bibinfo{author}{\bibfnamefont{D.}~\bibnamefont{Gu{\'e}ry-Odelin}},
 \bibinfo{journal}{Opt. Lett.} \textbf{\bibinfo{volume}{32}},
 \bibinfo{pages}{3143} (\bibinfo{year}{2007}).

 \bibitem[{\citenamefont{Gibble}(2009)\citenamefont{Gibble}}]{Gibble}
\bibinfo{author}{\bibfnamefont{K.}~\bibnamefont{Gibble}},
 \bibinfo{journal}{Phys. Rev. Lett.} \textbf{\bibinfo{volume}{103}},
 \bibinfo{pages}{113202} (\bibinfo{year}{2009}).


\end{thebibliography}

\end{document}